%% file: main.tex
\documentclass{article}
\usepackage{spconf,amsmath,graphicx}
\usepackage{float}
\usepackage{tikz}
\usepackage{amsfonts}
\usepackage{csquotes}
\usetikzlibrary{fit, backgrounds}
\usetikzlibrary{positioning, shapes.geometric, arrows.meta, shadows.blur}
\usetikzlibrary{positioning, shapes, arrows.meta, decorations.markings, decorations.pathreplacing, calc}
\usepackage[pagebackref,breaklinks,colorlinks]{hyperref}

\tikzstyle{block} = [rectangle, draw, fill=blue!20, text width=6em, text centered, minimum height=4em]
\tikzstyle{line} = [draw, -Latex]
\tikzstyle{sum} = [draw, circle, node distance=1cm]

\def\x{{\mathbf x}}

\title{Attention Down-Sampling Transformer, Relative Ranking and Self-Consistency for Blind Image Quality Assessment}
\name{Author(s) Name(s)}
\address{Author Affiliation(s)}

\name{Mohammed Alsaafin$^a$, Musab Alsheikh$^b$, Saeed Anwar$^{c,d}$, Muhammad Usman$^{c,d}$
}

\address{
    \tabular{c}
		King Fahd University of Petroleum and Minerals, Dhahran, 31261, Saudi Arabia\\
           $^a$ Department of Industrial and Systems Engineering, $^b$ Department of Electrical Engineering, $^c$ Department of \\Information and Computer Science, $^d$ SDAIA-KFUPM Joint Research Center for Artificial Intelligence\\
           \small{\{g201072600,  g202114890, 
 saeed.anwar,  muhammad.usman\}@kfupm.edu.sa}
	\endtabular
	\hskip 0.5in
}

\begin{document}
%
\maketitle
\begin{abstract}
\input{sections/abs}

\end{abstract}

\input{sections/intro}
\input{sections/related}
\input{sections/methodology}
\input{sections/experiments}
\input{sections/conclusion}
\bibliographystyle{ieeetr}
\bibliography{refs}

\end{document}

%% file: sections/abs.tex
The no-reference image quality assessment is a challenging domain that addresses estimating image quality without the original reference. We introduce an improved mechanism to extract local and non-local information from images via different transformer encoders and CNNs. The utilization of Transformer encoders aims to mitigate locality bias and generate a non-local representation by sequentially processing CNN features, which inherently capture local visual structures. Establishing a stronger connection between subjective and objective assessments is achieved through sorting within batches of images based on relative distance information. A self-consistency approach to self-supervision is presented, explicitly addressing the degradation of no-reference image quality assessment (NR-IQA) models under equivariant transformations. Our approach ensures model robustness by maintaining consistency between an image and its horizontally flipped equivalent. Through empirical evaluation of five popular image quality assessment datasets, the proposed model outperforms alternative algorithms in the context of no-reference image quality assessment datasets, especially on smaller datasets. Codes are available at \href{https://github.com/mas94/ADTRS}{https://github.com/mas94/ADTRS}

%% file: sections/keywords.tex
No-Reference Image Quality Assessment, CNNs, Transformers, Self-Consistency, Relative Ranking

%% file: sections/intro.tex
\section{introduction}
\label{sec:intro}
Understanding image quality is essential for many applications; however, it may be difficult since we periodically need an ideal reference image. We can address this issue using Non-Reference Image Quality Assessment (NR-IQA). The objective is to develop techniques that can independently assess the image's quality without needing the original image.
The importance of NR-IQA arises from its wide range of applications, including surveillance systems~\cite{lu2022blind}, medical imaging~\cite{li2018blind}, content delivery networks~\cite{yue2019blind}, image \& video compression~\cite{hu2021blind}, etc. It is vital in these domains to assess quality without the original reference image. NR-IQA advances imaging technology and improves user experience. Existing NR-IQA methods focus on developing novel algorithms to handle the problem of evaluating image quality. Test Time Adaptation technique for Image Quality Assessment (TTAIQA)~\cite{roy2023test}, Quality-aware Pre-Trained (QPT)~\cite{zhao2023quality} models through self-supervised learning, the Language-Image Quality Evaluator (LIQE), the data-efficient image quality transformer (DEIQT)~\cite{zhang2023blind} represents strides in this field and many methods that leverage CNNs. However, shortcomings persist, particularly the limitation imposed by the scarcity of labeled data, hindering the effectiveness of deep learning models and capturing only local features via CNNs while disregarding the nonlocal features of the image that transformers can capture. Popular datasets like the largest NR-IQA dataset, FLIVE, fall short compared to those in other domains, impeding the robust training of NR-IQA models.

Our main contribution is to develop an enhanced NR-IQA model to elevate its performance based on established metrics by leveraging the transformer architecture to capture nonlocal features and CNNs to capture local features. We seek to assess the performance of our improved model against existing NR-IQA methods. We test our methods using the most popular image quality datasets like LIVE, TID2013, CSIQ, LIVE-C and KonIQ10K. We'll utilize metrics like Spearman's Rank-Order Correlation Coefficient (SRCC) and Pearson's Linear Correlation Coefficient (PLCC) to calculate how well our model works. The main goal of this study is to get better-performing models using famous performance metrics.

\begin{figure}[tbp]
\centering
\begin{tikzpicture}[
    block/.style={
        rectangle, 
        draw, 
        thick,
        text width=4em, 
        align=center, 
        minimum height=1em, 
        rounded corners,
        font=\small,
        top color=white,
        bottom color={rgb:cyan,4;green,2;blue,2},
        drop shadow={opacity=.5,shadow xshift=0.5ex,shadow yshift=-0.5ex}
    },
    line/.style={
        draw, 
        -Latex,
        thick
    },
    node distance=0.4cm and 0.15cm
    ]

    \node (img1) {\includegraphics[width=0.05\textwidth]{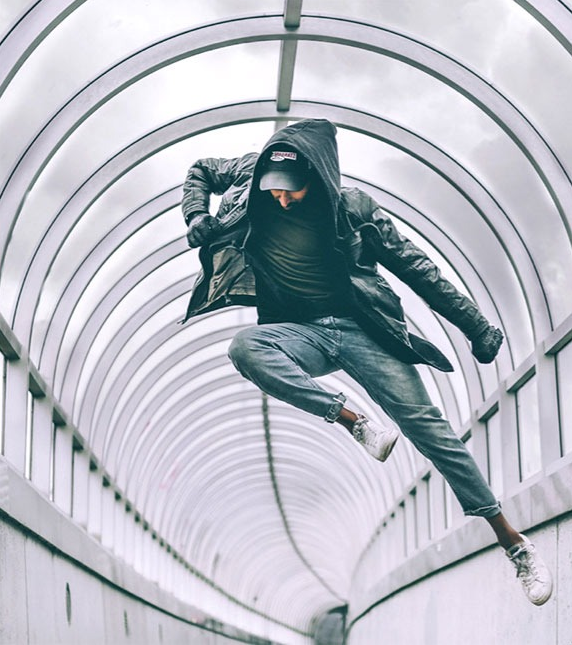}};
    \node[below=of img1] (img2) {\includegraphics[width=0.05\textwidth]{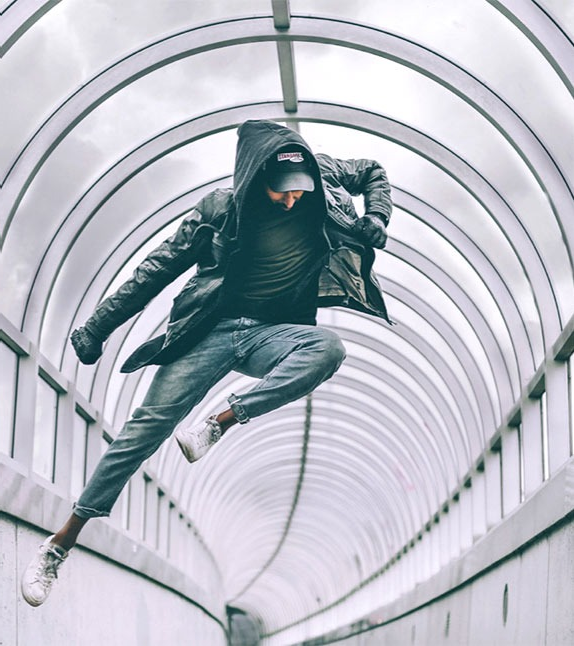}};

    \path let \p1 = ($(img1.south)!0.5!(img2.north)$) in node[block, fill=pink!30, right=2cm of img1, minimum width=1.5cm, minimum height=1.5cm, align=center] (model) at (\x1, \y1) {ADTRS\\Model};

    \draw[line, color={rgb:violet,2;blue,1;red,1}] (img1.east) -- ($(model.north west) - (0,0.59cm)$);
    \draw[line, color={rgb:violet,2;blue,1;red,1}] (img2.east) -- ($(model.south west)- (0,-0.59cm)$);
    \draw[line, color={rgb:orange,1;yellow,1;red,0.5}] ($(model.north east) - (0,0.85cm)$) -- ++(1,0) node[right] {Same Quality};

\end{tikzpicture}
\caption{Diagram illustrating the NR-IQA Model with inputs and outputs.}
\label{fig:nr_iqa_model}
\end{figure}
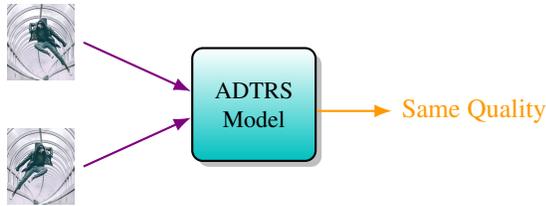

%% file: sections/related.tex
\section{related Works}
\label{sec:related}
Image quality assessment, or IQA, is an integral part of computer vision and image processing and is widely utilized in social networking, online content sharing, and photography. This review is divided into two parts. The first discusses the recently released papers and focuses on developing creative methods for assessing NR-IRQA. The second part focuses on papers that use convolutional neural networks (CNNs) and transformers for quality assessment.

Zhao~et~al.~\cite{zhao2023quality} provide a unique method for blind image assessment that uses self-supervised learning to get over the absence of labeled data. They propose a customized pretext task and a quality-aware contrastive loss, expanding image distortion simulations to enhance NR-IQA. Their Quality-conscious Pre-trained models beat current techniques on several BIQA benchmark datasets, demonstrating increased sensitivity for image quality. Distribution changes between training and testing situations in blind IQA significantly impact inference performance. Based on this, Subhadeep~et~al.~\cite{roy2023test} propose test-time adaptation by incorporating quality-relevant auxiliary activities at the batch and sample levels. Employing even a small batch of test data substantially improves model performance, surpassing existing SOTA approaches. 

A multitask approach~\cite{zhang2023blind} for blind image quality evaluation, leveraging auxiliary information from other tasks and automating model parameter sharing and weighting for loss functions, benefiting from scene categorization and distortion type identification tasks. The mentioned method outperforms existing approaches across multiple IQA datasets, enhancing resilience and aligning better with quality annotations. It also offers insights into the creation of next-generation NR-IQA models. 

Recently, a distinctive Mixture of Expert~\cite{saha2023re} introduces blind image quality assessment. This approach trains two separate encoders in an unsupervised setting to capture high-level content and low-level image characteristics. A linear regression model is developed to evaluate image quality by leveraging the synergy between these features. The authors showcase their technique's superiority over others across various extensive IQA databases, adeptly handling genuine and synthetic distortions. They highlight the significant impact of content-aware image representations, especially in an unsupervised context, on enhancing non-reference IQA performance.

The introduction of transformers~\cite{vaswani2017attention} as novel network architecture defines a novel attention mechanism to draw global dependencies between I/O while eliminating the need for recurrence and convolutions. Furthermore,  You~et~al. \cite{you2021transformer} were among the first papers to explore the transformer application in Image Quality (TRIQ) assessment. Building upon the original Transformer encoder in Vision Transformer, they presented a shallow Transformer encoder atop a convolutional neural network (CNN)-extracted feature map. Their architecture accommodates images of varying resolutions, employing adaptive positional embedding. They systematically evaluated different Transformer configurations across multiple publicly available image quality databases. Their research outcomes highlight the efficacy of the proposed TRIQ framework.

For blind image quality assessment, Zhang~et~al.~\cite{zhang2018blind} presented a unique deep bilinear model that can handle real and artificial distortions. Two CNNs designed for various distortion conditions make up their model. One CNN is pre-trained on a sizable dataset for visual distortion classification to handle synthetic distortions, using a previously trained CNN to classify images for real distortions. The two CNNs' features were bilinearly pooled to provide a cohesive interpretation for image quality estimate. The performance of the entire model is improved by fine-tuning it on target subject-rated datasets using a variation of stochastic gradient descent. Numerous tests confirm the model's exceptional performance on artificial and real image databases to approach the issue of no reference IQA as a learning-to-rank problem in which training utilizes the ranking data.

CNNs are good at translating images but struggle with rotations.  H-Nets~\cite{worrall2017harmonic}, an innovative solution to the problem of imagining translation and rotation impacting computer vision tasks differently, can accomplish 360-degree rotation equivariance and patch-wise translation using circular harmonics instead of standard CNN filters. Each receptive field patch receives the maximum responsiveness and orientation from this special construction. Deep feature mappings inside the network may encode complicated rotational invariants thanks to H-Nets' parameter-efficient representation with constant computational complexity. Their approach is easily incorporated into contemporary systems such as batch normalization and deep supervision. H-Nets compete well on benchmark problems and produce cutting-edge classification scores on rotated-MNIST, demonstrating their effectiveness. Their research indicates that the data remains vulnerable to equivariant transformations even with various augmentation techniques to increase CNN generalization. 

The field of blind image quality assessment has seen significant advancements recently, where each article addresses critical challenges in unique ways. The approaches leverage self-supervised learning, test-time adaptation, multitask learning, and unsupervised feature extraction to improve the precision and resilience of IQA models, making them better suited for real-world applications. Collectively, many studies contribute to the ongoing development of SOTA NR-IQA models, benefiting a wide range of industries and applications. Our primary goal in this paper centers around developing an advanced NR-IQA model that leverages transformers and CNNs to assess the quality of images with the primary aim of increasing its performance per established performance metrics. We intend to evaluate the ability of our enhanced model on five popular NR-IQA datasets by subjecting it to a thorough evaluation alongside existing NR-IQA approaches.

%% file: sections/methodology.tex
\section{Methodology}
\vspace{-1mm}
\label{sec:methodology}
In the proposed Attention Down-Sampling Transformer, Relative ranking and self-consistency abbreviated as ADTRS model for NR-IQA, we adopt the relative ranking and self-consistency mechanisms inspired by TReS~\cite{golestaneh2022no} but employ a completely different transformer architecture to evaluate the quality of images without reference standards. Our method begins with an input image from which a series of CNN layers extract crucial features representing varying complexities and scales. These features are then normalized and subjected to dropout to ensure the model's generalizability across diverse image sets.

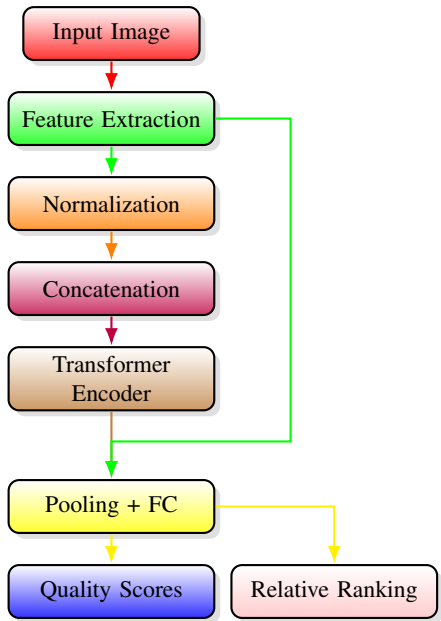
\begin{figure}[htbp]
    \centering
    \begin{tikzpicture}[
        block/.style={
            rectangle, 
            draw, 
            thick,
            text width=6em, 
            align=center, 
            minimum height=2em, 
            rounded corners,
            font=\small,
            top color=white,
            bottom color=#1!80,
            drop shadow={opacity=0.25, shadow xshift=0.5ex, shadow yshift=-0.5ex}
        },
        line/.style={
            draw, 
            -Latex,
            thick,
            color=#1,
            shorten >=0.2pt,
            shorten <=0.2pt
        },
        node distance=0.4cm and 0.2cm
        ]
        
        \node[block=red] (input) {Input Image};
        \node[block=green, below=of input, text width=2.5cm] (feature) {Feature Extraction};
        \node[block=orange, below=of feature, text width=2.5cm] (norm) {Normalization};
        \node[block=purple, below=of norm, text width=2.5cm] (concat) {Concatenation};
        \node[block=brown, below=of concat, text width=2.5cm] (transformer) {Transformer Encoder};
        \node[block=yellow, below=0.9cm of transformer, text width=2.5cm] (pooling) {Pooling + FC};
        \node[block=blue, below=of pooling, text width=2.5cm] (quality) {Quality Scores};
        \node[block=pink, right=of quality, text width=2.5cm] (ranking) {Relative Ranking};
        
        \draw[line=red] (input) -- (feature);
        \draw[line=green] (feature) -- (norm);
        \draw[line=orange] (norm) -- (concat);
        \draw[line=purple] (concat) -- (transformer);
        \draw[line=brown] (transformer) -- (pooling);
        \draw[line=yellow] (pooling) -- (quality);
        \draw[line=yellow] (pooling) -| (ranking);
        
        \coordinate (rightOfFeature) at ([xshift=1cm] feature.east);
        \coordinate (rightOfTransformer) at ([xshift=1cm] transformer.east);
        \coordinate (abovePooling) at ([yshift=0.5cm] pooling.north);
        
        \draw[line=green] (feature) -- (rightOfFeature) |- (rightOfTransformer) |- (abovePooling) -- (pooling);
        
    \end{tikzpicture}
    \caption{The basic building block of our the proposed ADTRS architecture.}
    \label{fig:nr_iqa_algorithm}
\end{figure}

As depicted in Figure~\ref{fig:nr_iqa_algorithm}, the workflow progresses by concatenating extracted features to create a cohesive feature set. The Transformer encoder employs self-attention to emphasize essential data elements, leveraging attention mechanisms to prioritize relevant information. Subsequently, the encoder's output is aggregated through a fully connected layer, facilitating dimensionality reduction and regression analysis. By incorporating self-consistency mechanisms, the model ensures the reliability of its predictions. In the final stage of the ADTRS model, it produces both absolute quality scores and relative rankings, enabling comprehensive image quality assessment and facilitating meaningful comparisons.

\subsection{Feature Extraction}

For an input image $I$ defined in the space $\mathbb{R}^{3 \times m \times n}$ with dimensions $m$ and $n$ symbolizing the width and height, respectively, the objective is to evaluate its perceptual QS. A CNN, represented by $f_\phi$ with learnable parameters $\phi$, is utilized to extract features $F_i$ from the $i^{t h}$ block, where $F_i \in \mathbb{R}^{b \times c_i \times m_i \times n_i}$ captures the feature maps with batch size $b$, and $c_i, m_i$, and $n_i$ denote the dimensions of the channels, width, and height of the extracted features, correspondingly.  

In neural network architectures, specifically in the context of deep learning and CNNs, a series of pre-processing steps are often applied to the extracted features from different layers. These steps include normalization, pooling, and dropout, each serving distinct purposes to enhance the network's performance and generalization. Normalization is employed to address variations in feature scales among different layers. Standardizing the features to have zero mean and unit variance or scaling them to a specific range ensures that they contribute more uniformly to the learning process. This step is crucial because features with different scales might dominate the learning, hindering the network's ability to converge effectively.

Pooling layers play a pivotal role in down-sampling the spatial dimensions of feature maps. By selecting the maximum or average values within specific regions, pooling reduces the computational complexity of the network, making it more efficient. Additionally, pooling contributes to the network's robustness by detecting invariant features and handling spatial variations in the input data. As a regularization technique, dropout addresses the risk of over-fitting by randomly deactivating a fraction of neurons during training. By preventing the network from relying too heavily on specific neurons, dropout encourages learning more robust features and improves the model's ability to generalize to unseen data. During testing, all neurons are reinstated to ensure the full utilization of the trained network. These pre-processing steps: normalization, pooling, and dropout—collectively constitute a comprehensive strategy for enhancing the efficiency, generalization, and robustness of neural networks, particularly in the complex tasks associated with deep learning architectures. (Eq.~\ref{eq:1}) is used to normalize the feature vector \( F_i \) using the Euclidean norm. The $L_2$ pooling is defined by (Eq.~\ref{eq:2}).

\begin{equation}
F_i = \frac{F_i}{\max\left(\|F_i\|_2, \epsilon\right)} \quad
\label{eq:1}
\end{equation}

\begin{equation}
P(x) = \sqrt{g \ast (x \odot x)}
\label{eq:2}
\end{equation}
where \( \odot \) denotes the point-wise product, and the blurring kernel \( g(\cdot) \) is implemented via a Hamming window that approximately applies the Nyquist criterion.

The extracted features will be concatenated after going through the normalization , pooling and dropout layers. 

\subsection{Transformer Encoder}
We adopt the encoder architecture from the literature to process multi-scale features $F_i$ from the CNN~\cite{carion2020end}. These features are sequenced and fed into the Transformer encoder~\cite{li2023rethinking}, where a multi-head, multi-layer self-attention mechanism, depicted in Figure~\ref{fig:nr_iqa_algorithm}, is employed to model the dependencies across the feature maps. The Transformer's architecture can learn complicated feature relationships without any built-in inductive bias. 

\begin{figure}[tbp]
    \centering
    \begin{tikzpicture}[
    block/.style={
        rectangle, 
        draw, 
        thick,
        rounded corners,
        font=\small,
        align=center,
        top color=white,
        bottom color=#1!80!white,
        blur shadow={shadow blur steps=5}
    },
    line/.style={-Latex, thick, draw=#1},
    node distance=0.6cm and 0.25cm
]

    \coordinate (start) at (-0.75,-1);
    \coordinate (pstart) at (-0.75,-1.25);
    \coordinate (downsample) at (0,-1);
    \coordinate (convbn) at (-0.75,-2);
    \coordinate (pconvbn) at (-0.75,-2.5);
    \coordinate (pool) at (0.0,-3.5);
    \coordinate (conv1) at (-3,-4.5);
    \coordinate (conv2) at (1.5,-3.5);
    \coordinate (q) at (0.5,-5.5);
    \coordinate (pq) at (0.5,-4.5);
    \coordinate (kt) at (-2,-5.5);
    \coordinate (v) at (-0.75,-5.5);
    \coordinate (attention) at (-0.75,-6.75);
    \coordinate (upsample) at (-0.75,-8);
    \coordinate (pend) at (-0.75,-8.5);
    \coordinate (end) at (-0.75,-9);
    
    \node[block=cyan, text width=2.2cm] at (convbn) (convbn_node) {Conv1x1 - BN};
    \node[block=cyan] at (conv1) (conv1_node) {Conv};
    \node[block=orange] at (kt) (kt_node) {K\textsuperscript{T}};
    \node[block=blue] at (pool) (pool_node) {Pool};
    \node[block=cyan] at (conv2) (conv2_node) {Conv};
    \node[block=green] at (q) (q_node) {Q};
    \node[block=violet] at (v) (v_node) {V};
    \node[block=yellow] at (attention) (attention_node) {Attention};
    \node[block=cyan, text width=2.2cm] at (upsample) (upsample_node) {Conv1x1 - BN};
    
    \draw[line=black] (start) -- (convbn_node);
    \draw[line=black] (pstart)-| (conv1_node);
    \draw[-=cyan] (convbn_node)-- (pconvbn);
    \draw[line=cyan] (pconvbn) -| (pool_node);
    \draw[line=cyan] (pconvbn) -| (conv2_node);
    \draw[line=cyan] (pconvbn) -| (kt_node);
    \draw[line=cyan] (pconvbn) -| (v_node);
    \draw[line=black] (conv1_node) |- (pend);
    \draw[-=blue] (pool_node) |- (pq);
    \draw[-=red] (conv2_node) |- (pq);
    \draw[line=magenta] (pq) -- (q_node);
    \draw[line=orange] (kt_node) |- (attention_node);
    \draw[line=green] (q_node) |- (attention_node);
    \draw[line=violet] (v_node) -- (attention_node);
    \draw[line=yellow] (attention_node) -- (upsample_node);
    \draw[line=cyan] (upsample_node) -- (end);
    
\end{tikzpicture}
\caption{Dual-Path downsampling transformer encoder~\cite{li2023rethinking} adopted in our proposed method.}
\label{fig:transformer}
\end{figure}
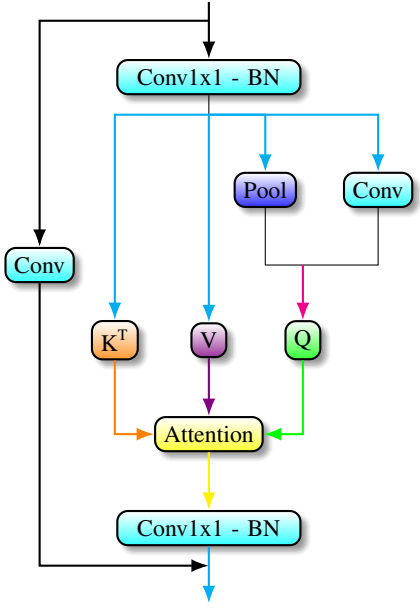

At the heart of our model lies the Transformer Encoder Layer paired with the Self-attention mechanism, essential for analyzing image characteristics through sophisticated spatial recognition. Initially, the model processes image features that distill the essence of the visual input and integrates positional encoding to inject spatial context into these features. Following this, the multi-head self-attention framework comes into play, dissecting and assessing input segments by generating attention scores through queries, keys, and values, all created via adaptive linear transformations.

After the self-attention phase, the outputs are fused and normalized in the Add \& Norm step, a measure that stabilizes the learning process and embeds residual connections, vital for the architecture's depth and efficacy. A subsequent Feed-Forward Network (FFN) executes additional linear transformations, punctuated by ReLU activation, to polish the feature set further. The depth of this encoding process is represented by 'Nx', reflecting the iterations of the transformation sequence. The depicted data flow in Figure~\ref{fig:transformer}, especially the loopback arrows, emphasizes the Transformer's residual connections, ensuring a seamless and continuous information stream within the model's architecture.

\begin{figure}[tbp]
    \centering
    \begin{tikzpicture}[
        block/.style={
            rectangle, 
            draw, 
            thick,
            rounded corners,
            font=\small,
            top color=white,
            bottom color=#1!80!white, 
            blur shadow={shadow blur steps=5}
        },
        operator/.style={
            circle,
            draw,
            thick,
            inner sep=-1pt,
            minimum size=0.5cm,
            top color=white,
            bottom color=#1!80!white, 
            blur shadow={shadow blur steps=5}
        },
        line/.style={-Latex, thick, draw=#1},
        node distance=0.4cm and 0.15cm
        ]
        
        \node[block=red] (linear1) {Linear};
        \node[block=cyan, below=of linear1] (linear2) {Linear};
        \node[operator=orange, right=0.5cm of linear1] (dot) {$\times$}; 
        \node[block=violet, below=of linear2] (linear3) {Linear};
        \node[operator=green, right=0.5cm of linear3] (sum) {$+$}; 
        \node[block=yellow, right=0.5cm of dot] (softmax) {Softmax};
        
        \draw[line=red] (linear1.east) -- (dot.west);
        \draw[line=cyan] (linear2.east) -| (dot.south);
        \draw[line=orange] (dot.east) -- (softmax.west);
        \draw[line=yellow] (softmax.south) |- (sum.east);
        \draw[line=violet] (linear3.east) -- (sum.west);
        
    \end{tikzpicture}
    \caption{Schematic of the Self-Attention Mechanism.}
    \label{fig:self_attention_mechanism}
\end{figure}
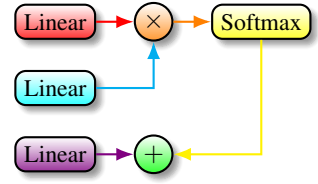

The self-attention mechanism within the Transformer model, as depicted in Figure~\ref{fig:self_attention_mechanism}, commences with dual linear blocks that reformulate the input data into intermediate states, integral for the ensuing attention calculations. This mechanism further employs the SoftMax function to normalize attention scores calculated from the dot products of queries and keys. This normalization facilitates a targeted distribution of attention across different data segments. Following this, the output from the SoftMax stage is combined with the outputs of a third linear block, symbolizing information integration within the attention process. This synthesis is not terminal but instead feeds back iteratively into the SoftMax stage, highlighting the recursive nature of the self-attention mechanism. This iterative loop, crucial for refining attention over successive cycles, is visually represented in Figure~\ref{fig:self_attention_mechanism}.

\textbf{Multi-Head Attention}: The multi-head attention mechanism, pivotal in our model, involves transforming input features into query, key, and value vectors. These vectors are then processed through the attention mechanism as shown in the following equations:

\begin{equation}
    \text{MultiHead}\left(Q^{\prime}, K^{\prime}, V^{\prime}\right)= \text{Concat}\left(\text {h}_1, \ldots, \text { h}_h\right) W^o,
    \label{eq:3}
\end{equation}
\begin{equation}
    \text {h}_i=\operatorname{Attention}\left(Q_i, K_i, V_i\right),
\end{equation}
\begin{equation}
    \text {Attention}(Q, K, V)=\operatorname{softmax}\left(\frac{Q K^T}{\sqrt{d_k}}\right) \odot V,
\end{equation}
where \( W^o \) is a projection matrix, \( Q \), \( K \), and \( V \) refer to the Query, Key, and Value matrices, respectively, and the softmax attention is normalized over the key dimension \( d_k \).

\subsection{Feature Fusion and Quality Estimation}
We utilize fully connected layers to combine the extracted features from convolutional and self-attention mechanisms. This fusion is instrumental in predicting the image's quality (as visualized in Figure~\ref{fig:nr_iqa_algorithm}). Formulated as follows, the model is trained to minimize the regression loss as:

\begin{equation}
    \mathcal{L}_{\text {Q }, B}=\frac{1}{N} \sum_{i=1}^N\left\|q_i-s_i\right\|,
\label{eq:6}
\end{equation}
where $q_i$ represents the predicted quality score for the $i^{th}$ image, while $s_i$ represents ground truth quality score.

\subsection{Relative Ranking Incorporation}
The regression loss effectively handles quality prediction; it overlooks ranking and correlation among images. We aim to account for the relative ranking within batches, focusing on extreme cases due to computational constraints. In image batch $B$, $qa_{\text{max}}$, $qa'_{\text{max}}$, $qa_{\text{min}}$, and $qa'_{\text{min}}$ represent predicted qualities for the highest, second highest, lowest, and second lowest subjective quality scores, respectively. Utilizing triplet loss with $d(x, y) = |x - y|$, we aim for constraints like $d(qa_{\text{max}}, qa'_{\text{max}}) + \text{margin}_1 \leq d(qa_{\text{max}}, qa_{\text{min}})$. Similarly, we desire $d(qa_{\text{min}}, qa'_{\text{min}}) + \text{margin}_2 \leq d(qa_{\text{max}}, qa_{\text{min}})$. Empirically selecting margin values is challenging due to dataset variations. For perfect predictions, $\text{margin}_1$ is bounded by $sqa'_{\text{max}} - sqa_{\text{min}}$, serving as an upper-bound during training, where $\text{margin}_1 = sqa'_{\text{max}} - sqa_{\text{min}}$, where $sqa'_{\text{max}}$  signifies the subjective quality score associated with the image having the predicted quality score $qa'_{\text{max}}$. We can do a similar process for $\text{margin}_2$ which is bounded by $sqa_{\text{max}} - sqa'_{\text{min}}$.

\begin{equation}
    \begin{aligned}
& \mathcal{L}_{\text {RR} \text {  }, B}= \\
& \mathcal{L}_{\text {triplet }}\left(qa_{\text {max }}, qa_{\text {max }}^{\prime}, qa_{\text {min }}\right)+\mathcal{L}_{\text {triplet }}\left(qa_{\text {min }}, qa_{\text {min }}^{\prime}, qa_{\text {max }}\right) \\
& =\max \left\{0, d\left(qa_{\text {max }}, qa_{\text {max }}^{\prime}\right)-d\left(qa_{\text {max }}, qa_{\text {min }}\right)+\text {margin}_1\right\} \\
& +\max \left\{0, d\left(qa_{\text {min }}^{\prime}, qa_{\text {min }}\right)-d\left(qa_{\text {max }}, qa_{\text {min }}\right)+\text {margin}_2\right\}
\end{aligned}
\label{eq:7}
\end{equation}

\begin{table}[tbp]
    \small
    \centering
        \caption{Summary of IQA datasets where \enquote{Dist.} stands for Distortion, and \enquote{\#} is for number.}
\begin{tabular}{lccc}
\hline Databases & \begin{tabular}{c} 
\# of Dist. \\
Images
\end{tabular} & \begin{tabular}{c} 
\# of Dist. \\
Types
\end{tabular} & \begin{tabular}{c} 
Dist. \\
Type
\end{tabular} \\
\hline LIVE & 799 & 5 & Synthetic \\
CSIQ & 866 & 6 & Synthetic \\
LIVE-C & 1.162 & - & Real \\
TID2013 & 3,000 & 24 & Synthetic \\
KonIQ10K & 10,073 & - & Real \\
\hline
\end{tabular}
\label{tab:1}
\vspace{-4mm}
\end{table}

\vspace{-4mm}
\subsection{Self-Consistency Mechanism}
For the last part of the methodology, we advocate leveraging the model's uncertainty in both the original input image and its equivariant transformation during the training process. To enhance the robustness of the model, we exploit self-consistency by establishing a self-supervisory signal between each image and its equivariant transformation. For a given input \( I \), denote the output logits from the Convolutional and Transformer layers as \( \zeta_{\epsilon, \text{conv}}(I) \) and \( \zeta_{\psi, \text{atten}}(I) \), respectively, where \( \zeta_{\epsilon, \text{conv}} \) and \( \zeta_{\psi, \text{atten}} \) represent the CNN and Transformer with learnable parameters \( \epsilon \) and \( \psi \), respectively. Our model utilizes these outputs to predict image quality. Given that human subjective scores remain consistent for the horizontally flipped version of the input image, we anticipate \( \zeta_{\epsilon, \text{conv}}(I) = \zeta_{\epsilon, \text{conv}}(\tau(I)) \) and \( \zeta_{\psi, \text{atten}}(I) = \zeta_{\psi, \text{atten}}(\tau(I)) \), where \( \tau \) signifies the horizontal flipping transformation. Consequently, by incorporating our consistency loss, the network learns to fortify its representation learning autonomously, eliminating the need for additional labels or external supervision. We aim to minimize the self-consistency loss

\begin{equation}
    \begin{aligned}
        & \mathcal{L}_{\text{SC}}=  \left\|\zeta_{\epsilon, \text{conv}}(I) - \zeta_{\epsilon, \text{conv}}(\tau(I))\right\| + \\
        & \left\|\zeta_{\psi, \text{atten}}(I) - \zeta_{\psi, \text{atten}}(\tau(I))\right\| + \theta_1\left\|\mathcal{L}_{\text{RR}, B} - \mathcal{L}_{\text{RR}, \tau(B)}\right\|,
    \end{aligned}
\label{eq:8}
\end{equation}
where $\tau(B)$ signifies the equivariant transformation on image batch $B$.

\vspace{-4mm}
\subsection{Composite Loss Function}
The overall training process involves the minimization of a composite loss function (Eq.~\ref{eq:9}), which encompasses quality loss (Eq.~\ref{eq:6}), relative ranking loss (Eq.~\ref{eq:7}), and self-consistency loss (Eq.~\ref{eq:8}). These losses are balanced by coefficients $\theta_1$, $\theta_2$, and $\theta_3$ to optimize the training outcome effectively. Our proposed model aims to provide a comprehensive and reliable NR-IQA solution by incorporating these elements. The effectiveness of this approach is demonstrated in the experimental results section.
\begin{equation}
\begin{split}
    \mathcal{L}_{\text {CLF }}  =\mathcal{L}_{\text {Q}}+\theta_2 \mathcal{L}_{\text {RR }}+ 
 \theta_3 \mathcal{L}_{\text {SC }}  
\end{split}
\label{eq:9}
\end{equation}

%% file: sections/experiments.tex
\begin{table*}[tbp]
    \small
    \centering
    \caption{Comparison of ADTRS and No-Reference State-of-the-Art Algorithms.}
    \begin{tabular}{l|cc|cc|cc|cc|cc}
\hline & \multicolumn{2}{|c|}{ LIVE } & \multicolumn{2}{c}{ CSIQ } & \multicolumn{2}{|c|}{ TID2013 } & \multicolumn{2}{|c|}{ LIVE-C } & \multicolumn{2}{|c}{ KonIQ10K }\\
\cline { 2 - 11 } & PLCC & SROCC & PLCC & SROCC & PLCC & SROCC & PLCC & SROCC & PLCC & SROCC\\
\hline DBCNN \cite{zhang2018blind} & 0.971 & 0.968 & 0.959 & $\mathbf{0 . 9 4 6}$ & 0.865 & 0.816 & 0.869 & $\mathbf{0 . 8 6 9}$ & 0.884 & 0.875\\
TReS \cite{golestaneh2022no} & 0.968 & 0.969 & 0.942 & 0.922 & 0.883 &  0.863 & 0.877 & 0.846 & $\mathbf{0 . 9 2 8}$ & $\mathbf{0 . 9 1 5}$\\
HFD \cite{wu2017hierarchical} & 0.971 & 0.951 & 0.890 & 0.842 & 0.681 & 0.764 & - & - & - & -\\
PQR \cite{zeng2017probabilistic} & 0.971 & 0.965 & 0.901 & 0.873 & 0.864 & 0.849 & 0.836 & 0.808 & - &-\\
DIIV INE \cite{saad2012blind} & 0.908 & 0.892 & 0.776 & 0.804 & 0.567 & 0.643 & 0.591 & 0.588 & 0.558 & 0.546\\
BRISQUE \cite{mittal2012no} & 0.944 & 0.929 & 0.748 & 0.812 & 0.571 & 0.626 & 0.629 & 0.629  & 0.685 & 0.681\\
ILNIQE \cite{zhang2015feature} & 0.906 & 0.902 & 0.865 & 0.822 & 0.648 & 0.521 & 0.508 & 0.508  & 0.537 & 0.523\\
BIECON \cite{kim2016fully} & 0.961 & 0.958 & 0.823 & 0.815 & 0.762 & 0.717 & 0.613 & 0.613 & 0.654 & 0.651\\
MEON \cite{ma2017end} & 0.955 & 0.951 & 0.864 & 0.852 & 0.824 & 0.808 & 0.710 & 0.697 & 0.628 & 0.611\\
WaDIQaM \cite{bosse2017deep} & 0.955 & 0.960 & 0.844 & 0.852 & 0.855 & 0.835 & 0.671 & 0.682 & 0.807 & 0.804\\
TIQA \cite{gao2015learning} & 0.965 & 0.949 & 0.838 & 0.825 & 0.858 & 0.846 & 0.861 &  0.845 & 0.903 &  0.892\\
MetaIQA \cite{zhu2020metaiqa} & 0.959 & 0.960 & 0.908 & 0.899 & 0.868 & 0.856 & 0.802 & 0.835 & 0.856 & 0.887\\
P2P-BM \cite{ying2020patches} & 0.958 & 0.959 & 0.902 & 0.899 & 0.856 & 0.862 & 0.842 & 0.844 & 0.885 & 0.872\\
HyperIQA \cite{su2020blindly} & 0.966 & 0.962 & 0.942 & 0.923 & 0.858 & 0.840 & $\mathbf{0 . 8 8 2}$ & 0.859 & 0.917 & 0.906\\
\hline
ADTRS (Ours)  & $\mathbf{0 . 9 7 2}$ & $\mathbf{0 . 9 7 0}$ & $\mathbf{0 . 9 6 0}$ & 0.943 & $\mathbf{0 . 8 9 7}$ &  $\mathbf{0 . 8 7 8}$ & 0.864 & 0.836 & 0.918 & 0.905\\
\hline
\end{tabular}
    \label{tab:2}
\end{table*}
\vspace{-5mm}
\section{Experiments}
\label{sec:experiments}
We assess our proposed ADTRS model's performance on five widely recognized IQA datasets, displayed in Table~\ref{tab:1} (among the distortions, three were synthetically generated while two occurred authentically). We utilize two standard performance metrics, PLCC and SROCC, among various metrics available in IQA evaluation. PLCC (Pearson Linear Correlation Coefficient) evaluates the correlation between algorithmic results and human eye subjective scores, reflecting the algorithm's accuracy. On the other hand, SROCC (Spearman Rank-Ordered Correlation Coefficient) measures the monotonicity of the algorithm's predictions. Both metrics range from 0 to 1, with higher values indicating superior performance.

\textbf{Implementation Details}:
We used an NVIDIA RTX 2060 GPU and PyTorch to train our model for training and testing. We augmented the horizontal and vertical dimensions of 138 randomly chosen patches, each measuring 224 by 224 pixels, from each image, by accepted IQA training protocols. The quality scores of the original image were carried over to these patches. With a weight decay of $5 \times 10^{-4}$ across a maximum of five epochs, we trained by minimizing the composite loss function over the training set, changing the learning rate from 2 $\times 10^{-5}$ and decreasing it by a factor of 5 after each epoch. A total of 138 patches, each measuring $224 \times 224$, were chosen randomly from the test picture during testing, and the final quality score was calculated by averaging their projected values. ResNet50~\cite{he2016deep} served as the CNN backbone for our model, which was seeded using Imagenet weights. We set the hyperparameters $\theta_1, \theta_2$, and $\theta_3$ to $0.5,0.05$, and 1 accordingly, using the Transformer architecture with 4 encoder layers, a hidden layer dimensionality of 64 $(d=16)$, and 16 heads $(h=16)$. All experiments were carried out with a consistent setup, adhering to NR-IQA standards. Datasets were randomly divided into $80 \%$ and $/ 20 \%$ train/test ratios.

\textbf{Performance Evaluation}: Table~\ref{tab:2} presents a comprehensive performance comparison based on PLCC and SROCC metrics across five standard image quality datasets. Our ADTRS consistently exhibits superior performance in both PLCC and SROCC evaluations compared to existing algorithms. Notably, among these SOTA NR-IQA algorithms, some incorporated CNNs while others did not. We chose the following algorithms since they were the most recent algorithms in \cite{golestaneh2022no} based on the same performance metrics. The bolded entries in Table 2 are the best-performing algorithms for each dataset. Drawing inspiration from TReS \cite{golestaneh2022no}, we adopted the relative ranking and self-consistency mechanism, making it equitable to compare our improved model (ADTRS) with it. As evident from Table 2, our model, employing a distinct transformer architecture, outperforms TReS and all other algorithms, particularly on smaller/synthetic datasets. Our model excels in Live, TID2013, and CSIQ's PLCC when compared against all different algorithms. Furthermore, our model performs exceptionally on real distorted and larger datasets, such as LIVE-C and KonIQ10K. In the KonIQ10K dataset, our model stands as the second-best performer, with TReS leading.

\textbf{Ablation Study \& Hyperparameter Tuning}: As previously mentioned, we employed ResNet50 as the experiments' primary backbone; smaller backbones offered faster processing speeds, yielding relatively inferior results. We also explored ResNet34 and ResNet18, but they produced significantly worse outcomes. Moreover, we experimented with varying sample patch sizes for training and testing hyperparameters, ranging from 16 to 138. The results represent the optimal outcomes from our extensive experiments, achieved with 138 sample patches. Additionally, we investigated enhancing the batch size from 8 to 16, which led to poorer results than using 8 batches. To fine-tune our model, we conducted a grid search across hyperparameters. This involved running the grid search for 5 epochs and adjusting parameters with significant performance impacts. For instance, we explored different configurations of encoder layers, such as 2, 4, 6, and 8 layers, along with varying values for other influential hyperparameters, e.g., training and testing sample patches. 

%% file: sections/conclusion.tex
\section{Conclusion}
\vspace{-2mm}
\label{sec:conclusion}
Our study presents an enhanced NR-IQA algorithm that efficiently merges CNNs and Transformer features, exploiting both local and non-local image characteristics for a comprehensive representation. We have incorporated a relative ranking loss function to capture essential ranking information among images, thereby augmenting the discriminative power of our model. By utilizing equivariant image transformations for self-supervision, we have bolstered the robustness of our approach. The performance of our method across five distinct IQA datasets underscores its robustness and adaptability. Our proposed algorithm outperforms all other algorithms on smaller and synthetic datasets while performing exceptionally well on larger datasets compared to different state-of-the-art algorithms. The results unequivocally establish the robustness and precision of our proposed method compared to the TReS model in accurately assessing image quality, highlighting its significant potential for diverse applications in image analysis and assessment.

\vspace{2mm}
\noindent
\textbf{Acknowledgement:}
The authors would like to acknowledge the support received from Saudi Data and AI Authority (SDAIA) and King Fahd University of Petroleum and Minerals (KFUPM) under SDAIA-KFUPM Joint Research Center for Artificial Intelligence Grant No. JRC-AI-RFP-11.